\documentclass[twocolumn,preprintnumbers,amsmath,amssymb]{revtex4}
\usepackage{graphicx}
\usepackage{dcolumn}
\usepackage{bm}

\newcommand{\be}{\begin{equation}}
\newcommand{\bel}[1]{\begin{equation}\label{#1}}
\newcommand{\ee}{\end{equation}}
\newcommand{\bea}{\begin{eqnarray}}
\newcommand{\ba}{\begin{array}}
\newcommand{\eea}{\end{eqnarray}}
\newcommand{\ea}{\end{array}}

\begin{document}
\title{Characteristics of the asymmetric simple exclusion process in the presence of \\ quenched spatial disorder}
\author{M. Ebrahim Foulaadvand $^{1,2}$, Sanaz Chaaboki $^{1}$, and Modjtaba Saalehi $^{1}$ }

\address{$^1$ Department of Physics, Zanjan University, P.O. Box
45196-313, Zanjan, Iran.\\
$^2$ Department of Nano-Sciences, Institute for studies in
theoretical Physics and Mathematics (IPM),\\
 P.O. Box 19395-5531, Tehran, Iran. }

\begin{abstract}

We investigate the effect of quenched spatial disordered hopping
rates on the characteristics of the asymmetric simple exclusion
process (ASEP) with open boundaries both numerically and by
extensive simulations. Disorder averages of the bulk density and
current are obtained in terms of various input and output rates.
We study the binary and uniform distributions of disorder. It is
verified that the effect of spatial inhomogeneity is generically
to enlarge the size of the maximal current phase. This is in
accordance with the mean field results obtained by Harris and
Stinchcombe \cite{stinchcombe}. Furthermore, we obtain the
dependence of the current and the bulk density on the
characteristics of the disorder distribution function. It is shown
that the impact of disorder crucially depends on the particle
input and out rates. In some situations, disorder can
constructively enhance the current.

\end{abstract}



\maketitle

\section{Introduction}

Transport processes in disordered media constitute an important
class of problems especially in the light of their relevance to
the modelling of a vast variety of phenomena in physics and many
interdisciplinary areas. A partial list of applications includes
transport phenomena in porous media, diffusion in biological
tissues and conduction through composite solids
\cite{sahimi,bunde,hughes}. It is a well-established fact the
disorder can strongly affect the transport characteristics of
equilibrium as well as out of equilibrium systems. Among various
non equilibrium systems, low dimensional driven lattice gases have
played an important role in describing the transport in many
physical, chemical and biological processes
\cite{ligget,zia,schutz1,css99,hinrischsen}. In particular, one
dimensional driven diffusive systems in the absence of disorder
have been extensively studied during the past two decades and at
present there exists a rich literature of results both analytic
and numeric \cite{schutz1}. Phase structures of these systems are
well known. It is well understood that non-equilibrium systems can
exhibit phase transitions in low dimensions. A model which has
played a paradigmatic role in out of equilibrium statistical
physics is the {\it asymmetric simple exclusion process} (ASEP)
\cite{mcdonald}. The model is amenable to exact analytical
solution \cite{derida1,domany,derida2}. Therefore it is a natural
and important question to investigate the effect of quenched
disorder on the phase structure of ASEP. Recently some efforts
and new strides have been made in the challenge between disorder,
interaction and drive. The exploration of the disordered ASEP
began with a single defective site in a periodic chain by
Janowsky and Lebowitz \cite{lebowitz1,lebowitz2}. They showed
that even one defective site can remarkably lead to global
effects on the system current and density profile. Evans solved
the ASEP with moving impurities where particle hopping rates were
chosen randomly from a distribution function \cite{evans1}. It
was shown that special distribution functions can give rise to a
new phase transition analogous to Bose condensation.
Subsequently, Tripathy and Barma \cite{barma1,barma2} considered
the ASEP on a ring with many defective sites. Their investigation
revealed the existence of phase segregation in a wide range of
global densities in the chain. In conjunction with the results of
the ASEP on a ring, an investigation of the disordered ASEP in an
open chain was introduced by Kolomeisky \cite{kolomeisky1}. He
showed that in some ranges of input and output rates, a single
defect in the bulk could affect the system properties on a global
scale. Recently a new wave of attention has been created on the
disordered
ASEP\cite{krug,kolwanker,derida3,stinchcombe,chou,shaw1,shaw2,lakatos1,santen,lakatos2}.
In particular, Chou and Lakatos have studied the effect of a few
defective sites in the open ASEP \cite{chou}. Their investigations
have revealed that generically the disorder's impact is highest
when the number of defects is very small. Increasing the number of
defects above a certain value has no further effect on the system
current. The question of the effect of a single defect in the ASEP
coupled with a 3D bulk reservoir with adsorption/desorption
kinetics was recently addressed by Frey \cite{frey}. Besides, some
time-dependent aspects of the disordered ASEP has been discussed
by Barma \cite{barma3}. Our goal in this paper is to deal in some
more depth with the problem of the disordered ASEP. Especially we
will focus on the role of the binary distribution function where
it provides the possibility of simultaneous study of both the
strength and the density of disorder throughout the chain. Via
extensive MC simulations, we show for the binary and uniform
distribution functions the generic impact of disorder is to
reduce the size of the low and high density phases. More
interestingly, we show in some circumstances, disorder can
constructively act in a manner to increase the system current.

\section{ Description of the Problem and numerical solution}

To keep the paper self-contained, let us first define the
disordered ASEP. Imagine a one dimensional stochastic process
defined on a discrete 1D lattice of length $L$. Each site can hold
at most one particle. We assign an integer valued number $s_i$ to
each site $i$ (see figure (1) ). If site $i$ is occupied, $s_i$=1.
If it is empty then $s_i$ is zero. The system configuration at
each time $t$ is characterized by specifying the occupation
numbers $s_i ~~ i=1,\cdots, L$. During an infinitesimal time $dt$
each particle can stochastically hop to its rightmost neighbouring
site provided the target site is empty. If the target site is
already occupied by another particle, the attempted movement is
rejected. The hopping takes place with a site dependent rate $p_i$
which is drawn from a given distribution function $f(p)$.

\begin{figure}
\centering
\includegraphics[width=7.5cm]{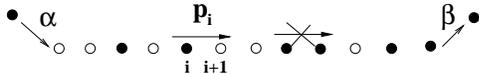}
\caption{ Asymmetric exclusion process with spatial disorder. } \label{fig:bz2}
\end{figure}

There is no spatial correlation between the set $p_i~~
i=1,\cdots,L$ and correspondingly the $p_i$'s can be regarded as
independent stochastic variables which are identically distributed
according to the site independent distribution function $f(p)$.
Denoting the averaged local density at site $i$ by $n_i$, one can
simply write the following rate equations for a particular
realisation of hopping rates $p_1, \cdots,p_L~ (i=2,\cdots,L-1)$:

\be \frac{d}{dt} \langle n_i \rangle= p_{i-1}\langle
n_{i-1}(1-n_i) \rangle - p_i\langle n_{i}(1-n_{i+1}) \rangle,
 \ee

\be \frac{d}{dt} \langle n_1 \rangle= \alpha(1-\langle n_1\rangle
)-p_1\langle n_1(1-n_2) \rangle, \ee \be \frac{d}{dt} \langle n_L
\rangle= p_{L-1} \langle n_{L-1}(1-n_L) \rangle- \beta \langle n_L
\rangle .\ee

No exact analytical solution exists for the above set of nonlinear
differential equations. Restricting ourselves to the stationary
state properties of the system, we set the left hand sides equal
to zero. We further simplify the equations by taking the
assumption of a mean field equation where the two-point functions
are replaced by the product of two one-point functions. This
assumption reduces the steady state equations into a set of
nonlinear algebraic equation with $L$ unknowns
$n_1,n_2,\cdots,n_L$.

\subsection{Numerical approach to mean field equations }

Even by employing the assumption of a mean field, we are not able
to solve the nonlinear algebraic equations. Therefore, we should
resort to numerical methods. We now outline a numerical approach
for solving the set of nonlinear equations. The approach is based
on the {\it shooting method} for solving boundary value problems.
To this end, we choose a trial $n_1$ denoted by $n_1^{tr}$ and
successively evaluate $n_2,\cdots,n_L$ through forward iteration.
The system current then turns out to be $\alpha (1-n_1^{tr})$.
Since the current should be equal for all sites, if the guessed
value of $n_1^{tr}$ was correct then the current evaluated from
the last site i.e., $\beta n_L$ would have the same amount $\alpha
(1-n_1^{tr})$ evaluated from the first site. To match these
currents, we gradually increase $n_1^{tr}$ from zero and evaluate
both currents from the first and last sites. Whenever these two
values become equal, then we have a solution. Note that in an
acceptable solution, all the densities $\langle n_1
\rangle,\cdots,\langle n_L \rangle$ should lie between $0$ and
$1$. We are interested in knowing the overall effect of disorder
on the transport characteristics of the ASEP. For given values of
$\alpha$ and $\beta$, we evaluate the current and density for
many samples of disordered chains and average over these samples.
We denote the sample averaged current and bulk density by
$\langle J \rangle $ and $\langle \rho \rangle$ respectively. For
obtaining a better insight, we have also executed extensive Monte
Carlo simulations. The disorder distributions we consider consist
of uniform and binary. More explicitly, the normalized uniform
distribution in the interval $[a,b]$ has the functional form
$f_1(p)=\frac{1}{b-a}$ with mean $\frac{a+b}{2}$ and variance
$\frac{(b-a)^2}{12}$. The binary distribution has the form
$f_2(p)=f \delta (p-p_1) +(1-f) \delta (p-p_2)$ where the binary
rates $p_1$ and $p_2$ and their probabilities $f$ and $1-f$ are
given. The mean value and the variance are $f p_1+(1-f) p_2$ and
$f p_1^2 +(1-f)p_2^2-[f p_1+(1-f) p_2]^2$ respectively. In the
subsequent sections, we show the result of simulation as well as
numerical solution of the mean field equations.

\section{Binary distribution function}

Let us first discuss the binary distribution of the quenched
disorder. Although in this type of distribution the defect
strength is allowed to take only an integer number of values (here
$2$), but even in this simple case one encounters some nontrivial
aspects which are worth investigating. Figure (2) depicts the
dependence of average current versus $\beta$ for some fixed
$\alpha$. The parameters of the binary distribution function is as
follows: $f=0.5,~p_1=0.8$ and $p_2=1.2$. The mean value of the
quenched hopping rate is fixed at $1$. The bulk density and the
current have been averaged over $1000$ disordered samples and the
system size is $200$.

\begin{figure}
\centering
\includegraphics[width=7.5cm]{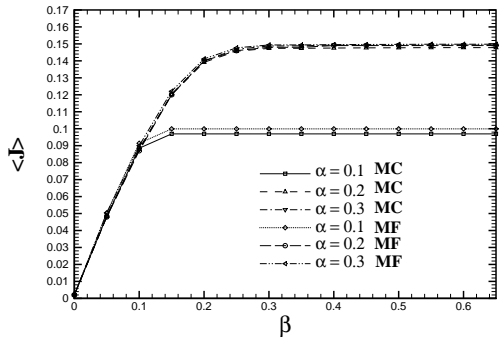}
\caption{ $\langle J \rangle$ versus $\beta$ for various input rates
$\alpha$. } \label{fig:bz2}
\end{figure}

One observes similar behaviour to the normal ASEP. Currents rise
up to a critical $\beta$ and then get saturated. The overall
effect of disorder is to reduce the value of the currents in each
phase. In the normal ASEP, the dependence of current on $\beta$ in
the high density (HD) phase is $\beta(1-\beta)$. Saturation of
current means that we are in the maximal current (MC) phase.
However, the current saturates at 0.15 which is less than the
value of the maximal current 0.25 in the normal ASEP. The reason
is due to the presence of defects which slow down the current.
Figure (3) exhibits the dependence of $ \langle J \rangle$ versus
the input rate $\alpha$.

\begin{figure}
\centering
\includegraphics[width=7cm]{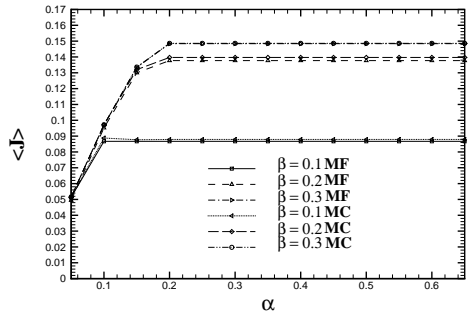}
\caption{ Disorder-averaged current vs $\alpha$ for various output
rates $\beta$. System size is 200. } \label{fig:bz2}
\end{figure}

The behaviour seen in the above graph is analogous to the normal
ASEP with the difference that the disorder has yielded to an
overall diminishing of $\langle J \rangle$. In the normal ASEP,
the dependence of $\langle J \rangle$ on $\alpha$ in the low
density phase is $\alpha(1-\alpha)$. We note that upon entering
the saturation regime i.e., the MC phase, the current value which
is 0.15, is less than that of the normal ASEP 0.25. The dependence
of bulk densities on $\alpha$ and $\beta$ are exhibited in figures
(4,5). Similar to current diagrams, the overall behaviour is
analogous to the normal ASEP. Here the effect of disorder is to
enhance the densities.

\begin{figure}
\centering
\includegraphics[width=7cm]{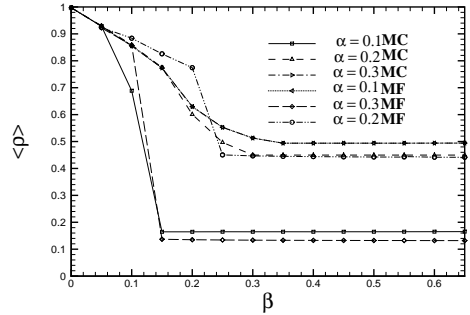}
\caption{ Disorder-averaged densities vs $\beta$ for various input
rates $\alpha$. System size is 200 and averaging has been executed
over 1000 disordered samples. } \label{fig:bz2}
\end{figure}

One observes the persistence of the first-order high to low
density transition. The saturation density is slightly above the
normal ASEP value $\alpha$ which is due to defects. For higher
$\alpha$ corresponding to the MC phase, the limiting density is
the same as the normal ASEP i.e., $\langle\rho\rangle=0.5$. This
indicates that the presence of defects does not alter the density
in the maximal current phase but reduces the current as discussed
earlier. The dependence of $\langle\rho\rangle$ on $\alpha$ is
shown in figure (5).

\begin{figure}
\centering
\includegraphics[width=7cm]{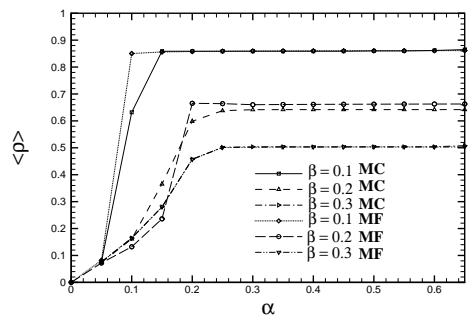}
\caption{ Disorder-averaged densities vs $\alpha$ for various output
rates $\beta$. System size is 200 and averaging has been executed
over 1000 disordered samples. } \label{fig:bz2}
\end{figure}

Similar to fig. (4), the density value in the saturation regime is
roughly 0.5 which is the same as the normal ASEP density value in
the MC phase. However, $\langle\rho\rangle$ differs from $1-\beta$
in the LD phase. We note that both high to low and low to high phase
transitions which are first order in the normal ASEP are replaced
with a smoother behaviour in the presence of disorder. We have
extensively performed Monte Carlo simulations for all ranges of
$\alpha$ and $\beta$. The simulation results confirm the existence
of three phases of LD, HD and maximal current. Furthermore, our
simulations show the growth of the maximal-current region and
shortening of the sizes of the LD and HD phases. These findings are
in agreement with the mean-field based conclusions of Harris and
Stinchcombe \cite{stinchcombe}. If one changes the parameters of the
binary distribution function, the overall picture remains
qualitatively the same as in the above diagrams. Nevertheless, the
quantitative values of both $\langle J \rangle$ and
$\langle\rho\rangle$ in the phases depend on the parameters of the
distribution functions. More concisely, currents and densities are
functional of $f(p)$. In the following figure, we exhibit the phase
diagram of disordered ASEP for some binary distribution functions.

\begin{figure}
\centering
\includegraphics[width=7cm]{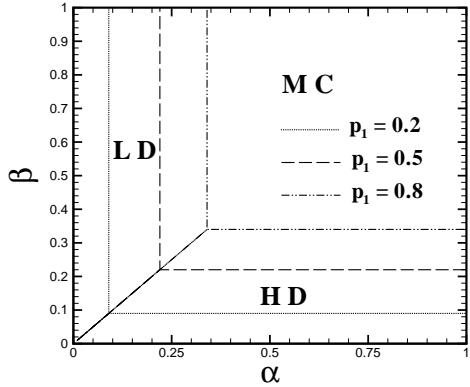}
\caption{ Phase diagram of the disordered ASEP for some binary
distributions. The distribution parameters are specified in the
figure. All the distributions have $f=0.5$ and $\langle p \rangle
=1$ but the variances are different. } \label{fig:bz2}
\end{figure}

We note that the size of the MC phase is an increasing function of
the variance of the distribution function. The reason is that
currents and densities are dominated by the number of defective
sites. If the variance of the distribution is large, then the
probability of finding sites with notably small hopping rates is
considerable and therefore $\langle J \rangle$ and $\langle \rho
\rangle$ are highly affected. The enlargement of the MC phase has
been reported for ASEP with a single defect in the bulk
\cite{kolomeisky1}. We recall from normal ASEP that the critical
values of the input and output rates are
$\alpha_c=\beta_c=\frac{p}{2}$ where $p$ is the hopping rate. In
principle, since $\alpha,\beta$ and $p$ are rates, they can vary
from zero to infinity. Therefore, it is possible to choose the
time unit such that $p$ scales to unity. In the disordered
version of the ASEP, one does not have a single hopping rate so it
would be better not to restrict ourselves to a particular time
unit. For the sake of comparison we write the values of the
density and current in the low density, high density and maximal
current phases of the normal ASEP:

\be \rho_{LD}=\frac{\alpha}{p},~\rho_{HD}=1-\frac{\beta}{p},~
\rho_{MC}=\frac{1}{2} . \ee Having in the mind that
$J=p\rho(1-\rho)$ we thus obtain: \be
J_{LD}=\alpha(1-\frac{\alpha}{p}). \ee \be
J_{HD}=\beta(1-\frac{\beta}{p}).   \ee \be
J_{MC}=\frac{p}{4}.                 \ee

Since in the binary distribution there are three parameters
namely $p_1, p_2$ and $f$, we have studied two distinguished
cases. First, we restrict ourselves to the condition $\langle p
\rangle=fp_1+(1-f)p_2=1$. This leaves only two free parameters.
In the second case, we impose the condition $p_2=1$ while $f$ and
$p_1$ are free to take arbitrary values in $[0,1]$. In the latter
case, the emphasis is on the role of defective sites $(p_1 <1)$
among normal sites $(p_2=1)$ whereas the former case allows
having fast hopping sites with rate $p_2>1$. Our simulations
showed that there is no significant differences between the
results of these two cases. Therefore in what follows, we only
exhibit the results for the case $p_2=1$ i.e., slow defective
sites among normal sites. Our first set of graphs (all obtained
via Monte Carlo simulations) illustrates the dependence of
$\langle J \rangle$  on $f$ for various $p_1$ in three sets of
input and output rates corresponding to low input-high output,
high input-low output and high input-high output.

\begin{figure}
\centering
\includegraphics[width=7cm]{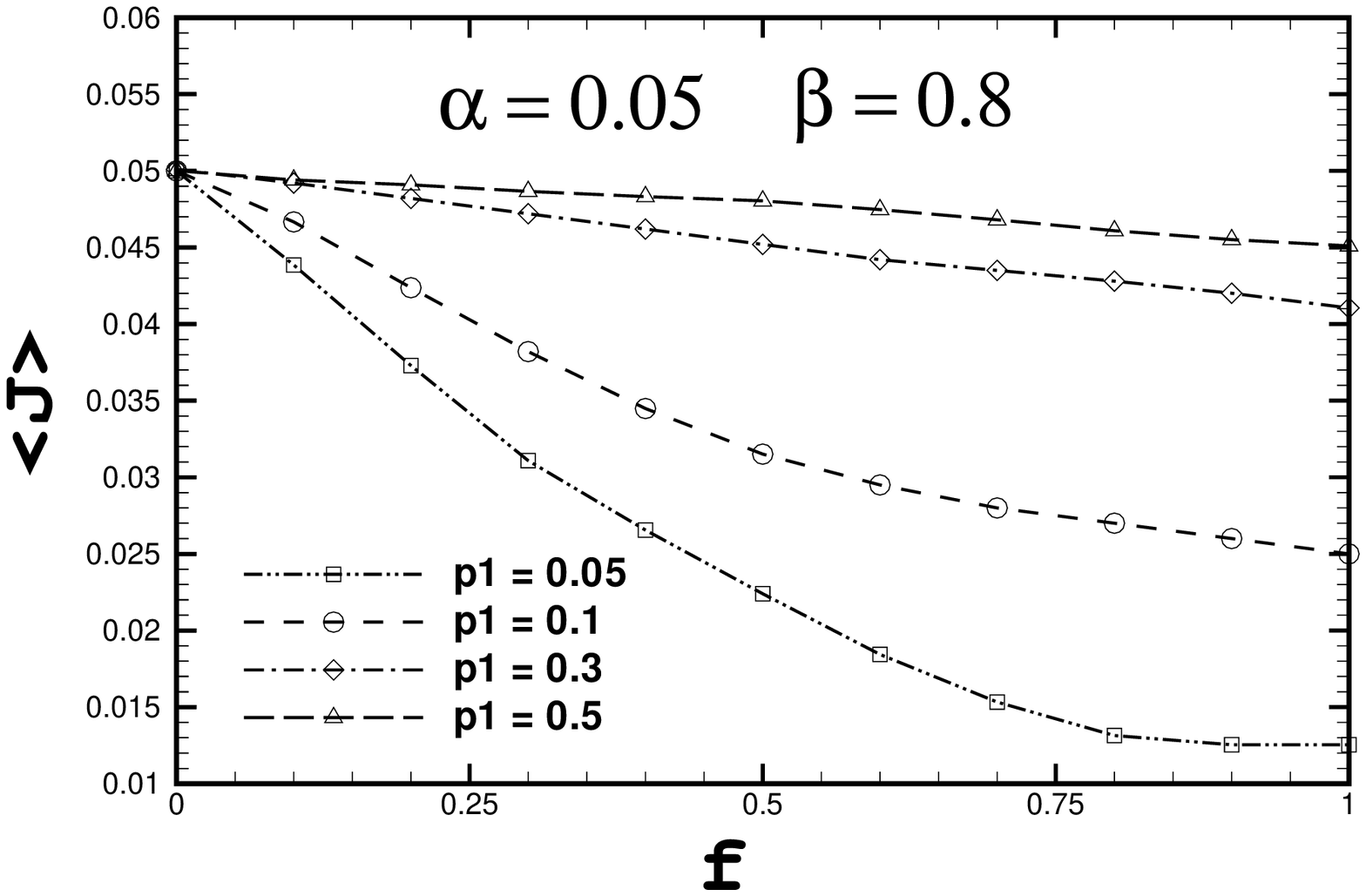}
\includegraphics[width=7cm]{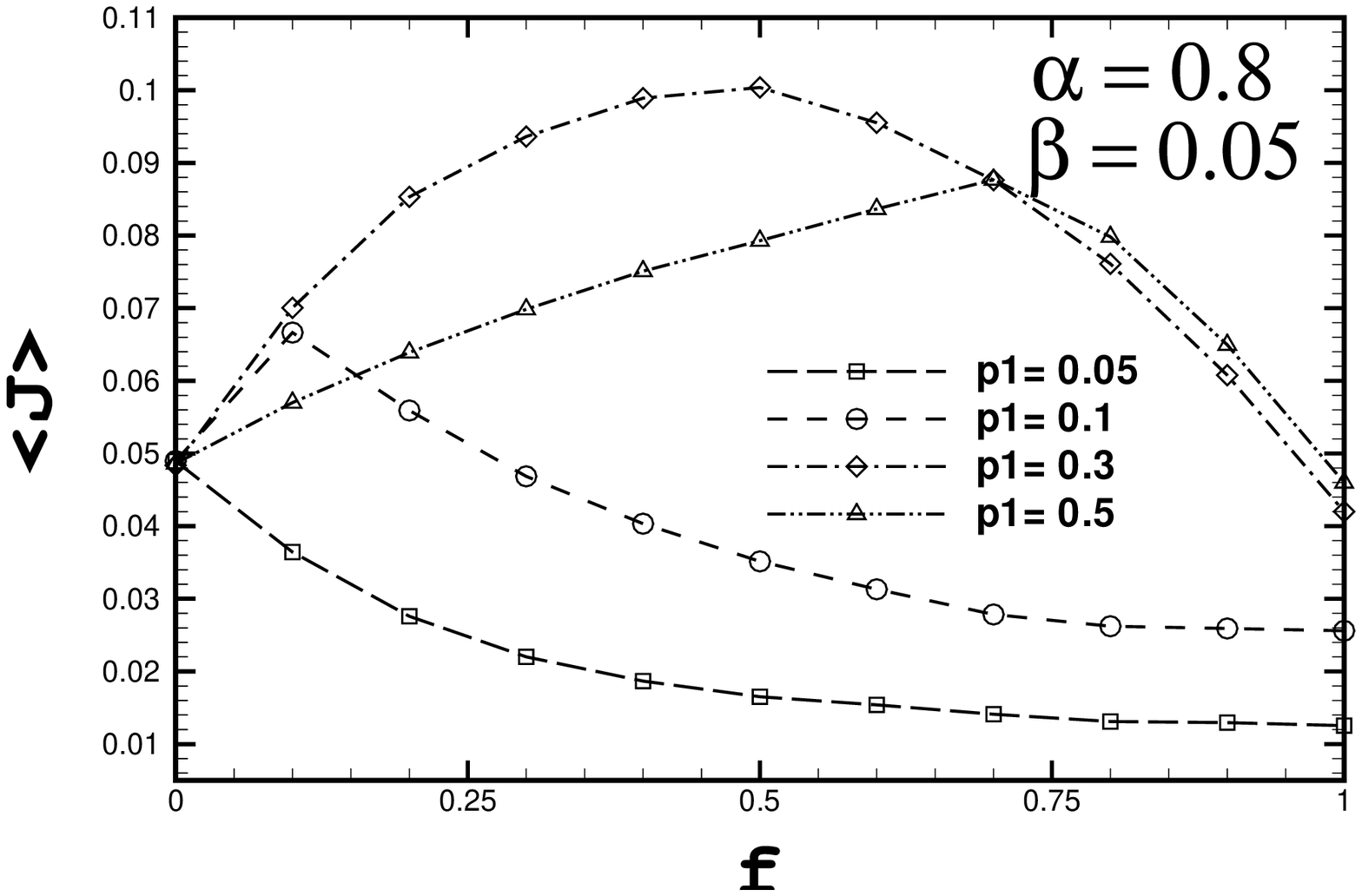}
\includegraphics[width=7cm]{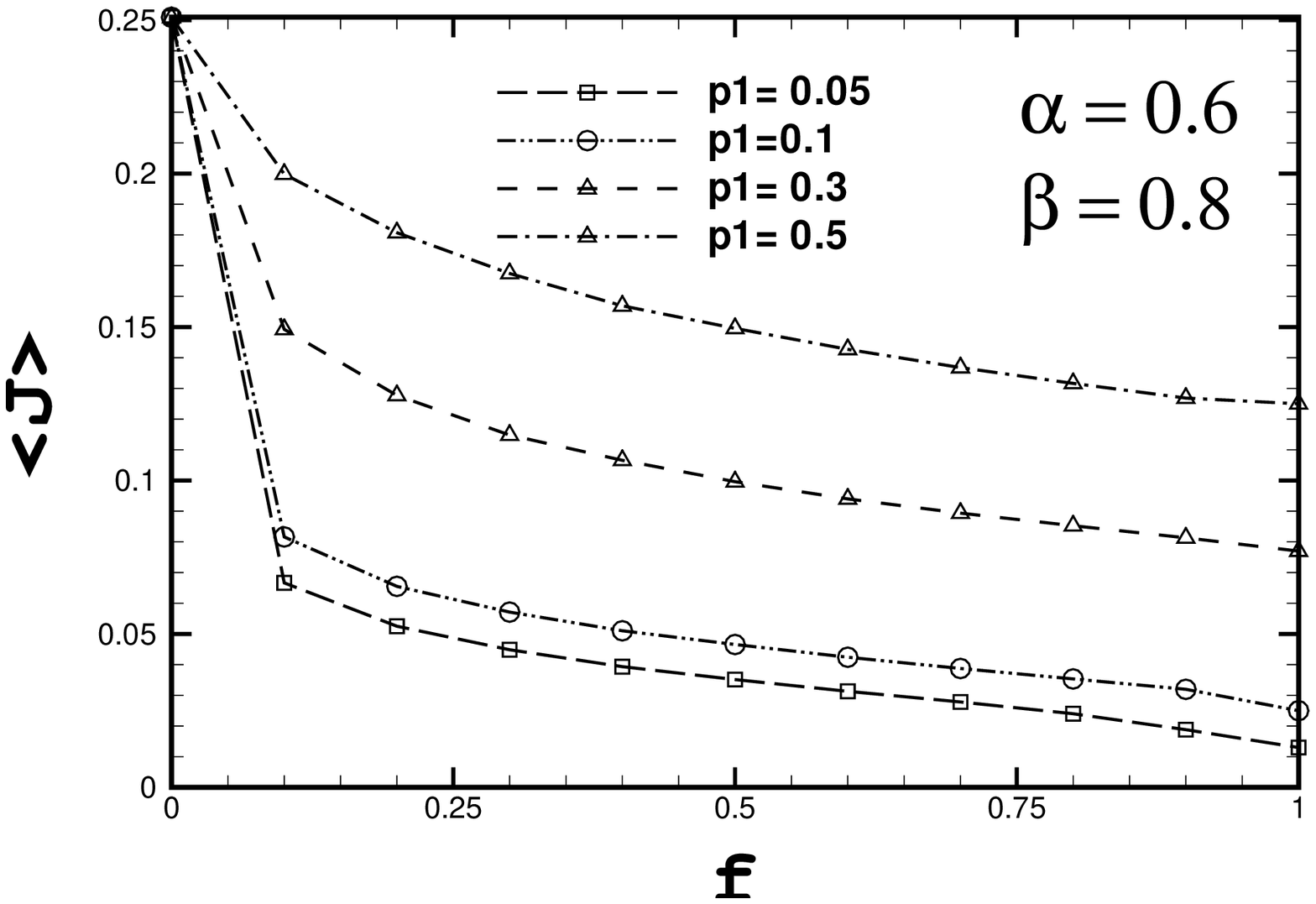}

\caption{ Figures~7-9: $\langle J \rangle$ vs $f$ for various $p_1$:
$\alpha=0.05,~\beta=0.8$ (top), $\alpha=0.8, ~\beta=0.05$ (middle)
and $\alpha=0.6,~\beta=0.8$ (bottom). System size is $L=300$. }
\label{fig:bz2}
\end{figure}

When $\alpha$ is small and $\beta$ is high (fig.7), the effect of
increasing $f$ is to reduce the current. Smaller values of $p_1$
exhibit a sharper decrease. This is natural since the bulk density
is low and therefore the system current is more sensitive to both
the number and the strength of the defects. The dependence of
$\langle J \rangle$ on $f$ changes qualitatively when one goes to
the situation characterized by high $\alpha$ and low $\beta$ (
fig.8 ). Here we are confronted with unexpected and novel
features. For $p_1$ less than $0.15$, the current appears as a
decreasing function of $f$ while for $p_1 > 0.1$ it increases up
to a maximum and then starts diminishing. Accordingly, the optimum
value of $f$ at which $\langle J \rangle$ is maximum is no longer
$f=0$ but rather a nonzero $f_{max}$. This implies that the
effect of disorder is to enhance the current which is a desirable
effect. The location of $f_{max}$ shifts towards higher values
when one increases $p_1$. For large $p_1$, $\langle J \rangle$
shows an increasing behaviour with a small slope. The slope tends
to zero when $p_1 \rightarrow 1$. For relatively high values of
$\alpha$ and $\beta$ (fig. 9), one still observes that the
dependence of current versus $f$ shows a decreasing character. In
this case, the disorder has the expected behaviour i.e.; the
higher the number of the impurities, the larger the decrease of
$\langle J \rangle $. However, the interesting point is the
abrupt change in the behaviour of the current reduction. For each
$p_1$, the current shows a rapid reduction up to a certain $f$
and then decreases very smoothly in a nonlinear fashion. This
marks the fact that impurities affect the system beyond a certain
relative frequency. These results are in agreement with
\cite{lakatos2}. To gain a deeper insight, it would be
instructive to look at the behaviour of $\langle \rho \rangle $
vs $f$. In low input and high output rates (fig.10), one observes
that for low $p_1$ the bulk density rises up to a maximum and
then decreases even below the value of normal ASEP. One might
naively think that increasing the density of defective sites
leads to an enhancement of the bulk density due to the formation
of high density regions behind them. However, the point is that
if the density of defective sites reaches a certain value, the
probability of finding defective sites in the vicinity of the
first site of the chain increases too.

\begin{figure}
\centering
\includegraphics[width=7cm]{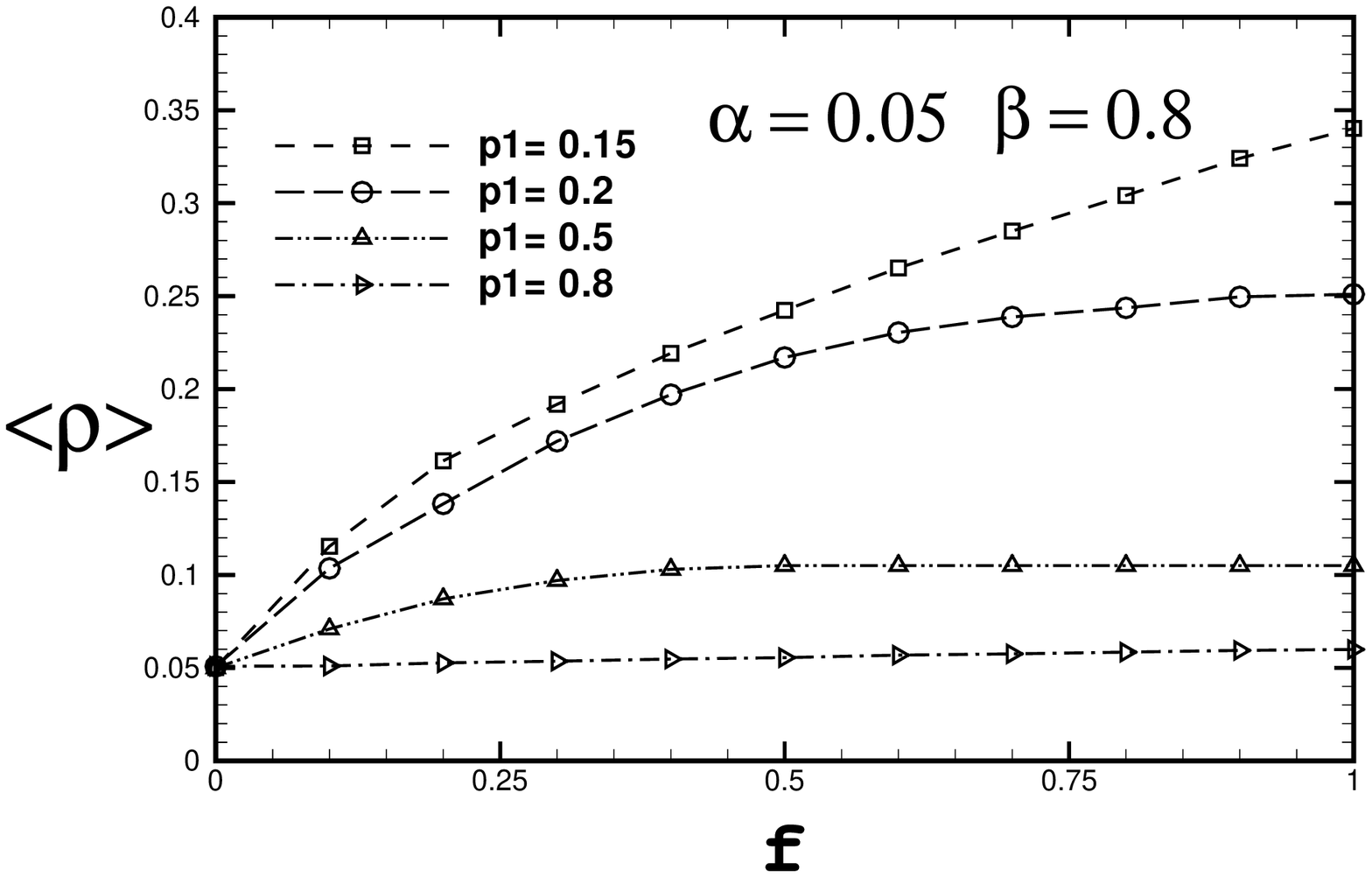}
\includegraphics[width=7cm]{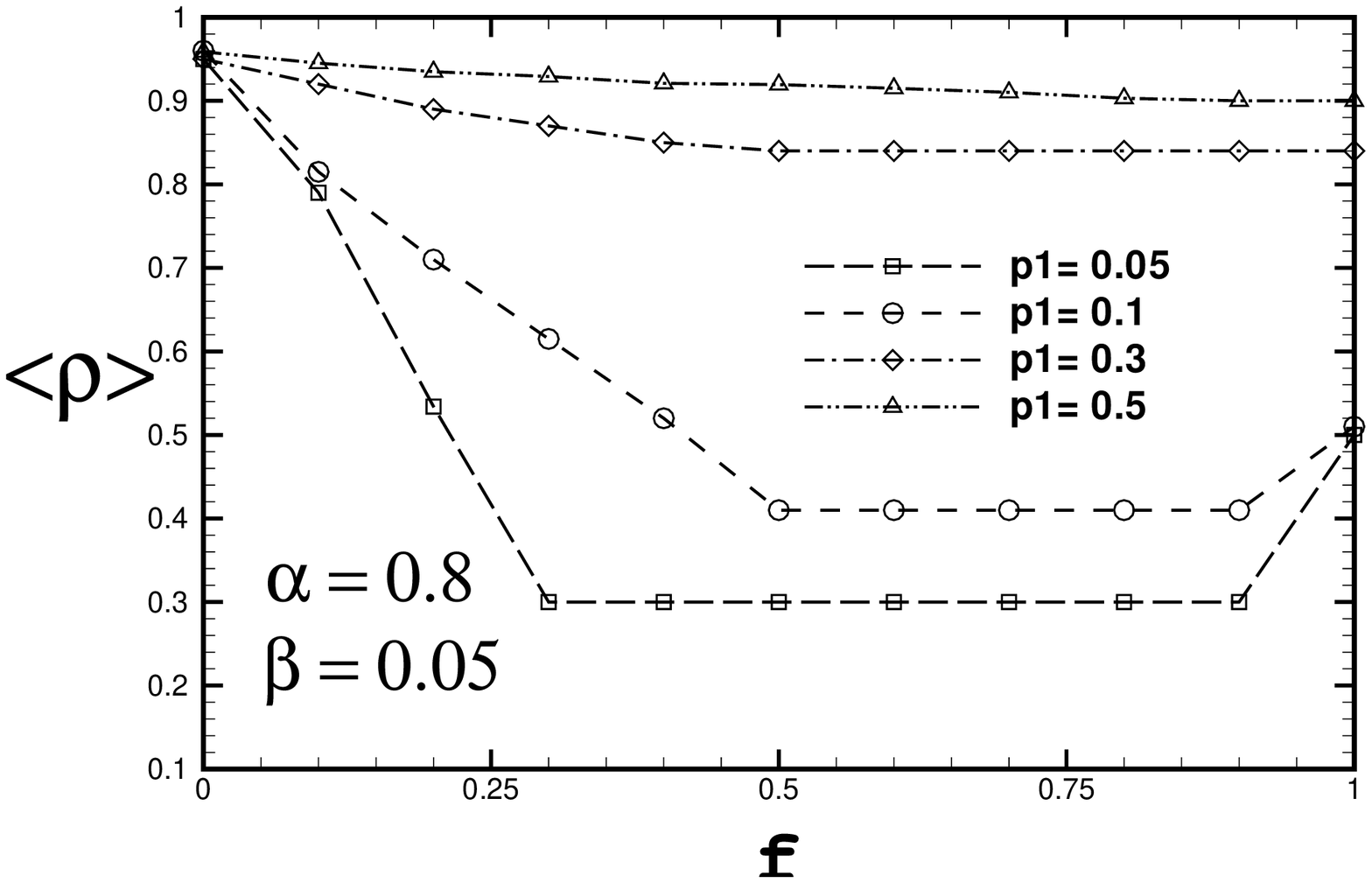}
\includegraphics[width=7cm]{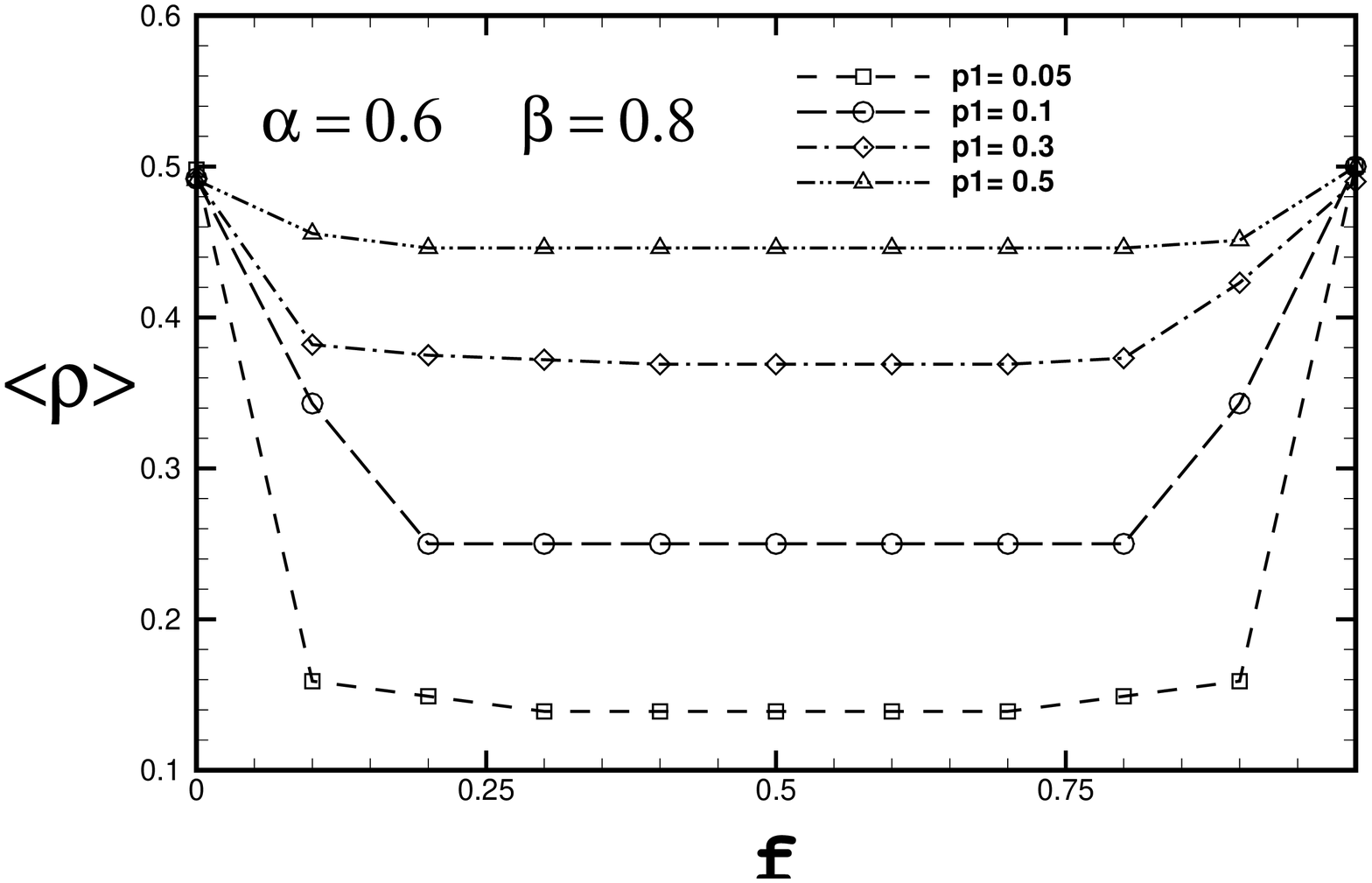}

\caption{ Figures~ 10-12: Bulk density dependence on $f$ for various
values of $p_1$: low density phase $\alpha=0.05, \beta=0.8$ (top),
high density phase $\alpha=0.8, \beta=0.05$ (middle) and maximal
current phase $\alpha=0.6, \beta=0.8$ (bottom). System size is
$L=300$. } \label{fig:bz2}
\end{figure}

This in turn gives rise to a blocking of the current of particles
in the chain bulk. As a result, a large portion of the bulk
remains almost in a LD regime which leads to a decreasing in the
bulk density throughout the chain. This scenario remains valid for
small $p_1$. For larger $p_1$, one observes the expected increase
of the bulk density upon increasing the density of defective
sites $f$. The reason is that once the defect strength is reduced
below a threshold, the formation of high density regions behind
these weak sites will be suppressed and therefore the particles
can more easily flow throughout the bulk. As a result of this
flow, enough particles can be found in the bulk. This increases
the number of local high regions behind defects which in turn
give rises to the enhancement of $\langle \rho \rangle$. In the
high $\alpha$-low $\beta$ regime (fig. 11) and for $p_1<0.2$ the
dependence of the bulk density on $f$ is sharply decreasing up to
a certain $f$ . Afterwards, $\langle \rho \rangle$ becomes
independent of $f$ and a lengthy plateau region forms. At
$f=0.9$, $\langle \rho \rangle$ shows a rather linear increase to
its asymptotic value $0.5$ in the MC phase. We note that when
$f=1$, all the sites are defective. For instance, in the case
$p_1=0.1$, the critical input and output rates are
$\alpha_c=\beta_c=\frac{p_1}{2}=0.05$. In this case $\alpha=0.8,
\beta=0.05$ lie in the MC phase and hence $\langle \rho \rangle$
approaches $0.5$ in the limit $f \rightarrow 1$. The reason is
that when the input is high and the output is low, impurities give
rise to phase segregation behind them \cite{barma2,kolomeisky1}.
The formation of macroscopic low density regions in front of them
leads to a sharp reduction of $\langle \rho \rangle$. For
$p_1>0.2$ the decrease of $\langle \rho \rangle$ becomes much
more smooth. The reason is that weaker defects are unable to
produce low enough density regions. The other interesting point
is that when the input rate is high, increasing the number of
defects will prevent a high inflow of particles, which is due to
the largeness of $\alpha$ and regulates the flow along the bulk.
The overall effect is to reduce $\langle \rho \rangle$ from high
$\alpha$ to much lower values. As we had already seen in the
current diagrams, in some values of $p_1$, this diminishing in
the bulk density is accompanied by the current increment as
exhibited in fig. (8). Now let us discuss the regime where
$\alpha$ and $\beta$ are both greater than $0.5$ (fig. 12). For
weak defect strength, the density is almost independent of $f$.
For $p_1$ less than 0.5, $\langle \rho \rangle$ shows a smoothly
decreasing dependence on $f$ until it becomes independent of $f$
and correspondingly a plateau region forms. The length of the
plateau is relatively large and becomes larger for smaller values
of $p_1$. Increasing $f$ beyond the plateau value, one again
encounters an increasing behaviour of $\langle \rho \rangle$ until
it reaches the normal ASEP value $\langle \rho \rangle=0.5$ in
the limit $f=1$ in which all sites have become defective. In order
to shed more light on our understanding, we now study the effect
of varying the disorder strength $p_1$ for fixed values of $f$.
Analogous to the previous studies, we first consider the current
which is shown in figures (13-15).

\begin{figure}
\centering
\includegraphics[width=7cm]{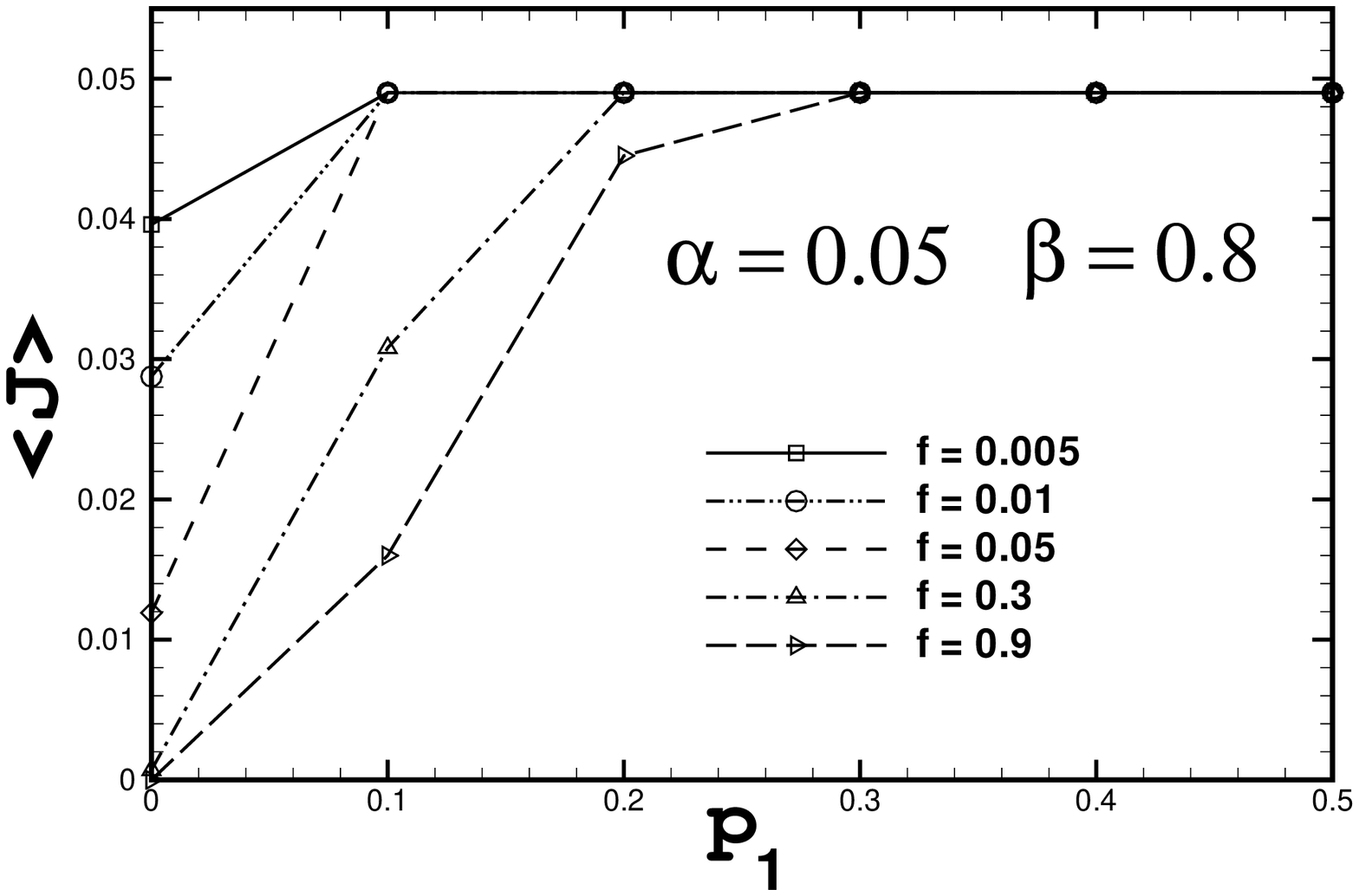}
\includegraphics[width=7cm]{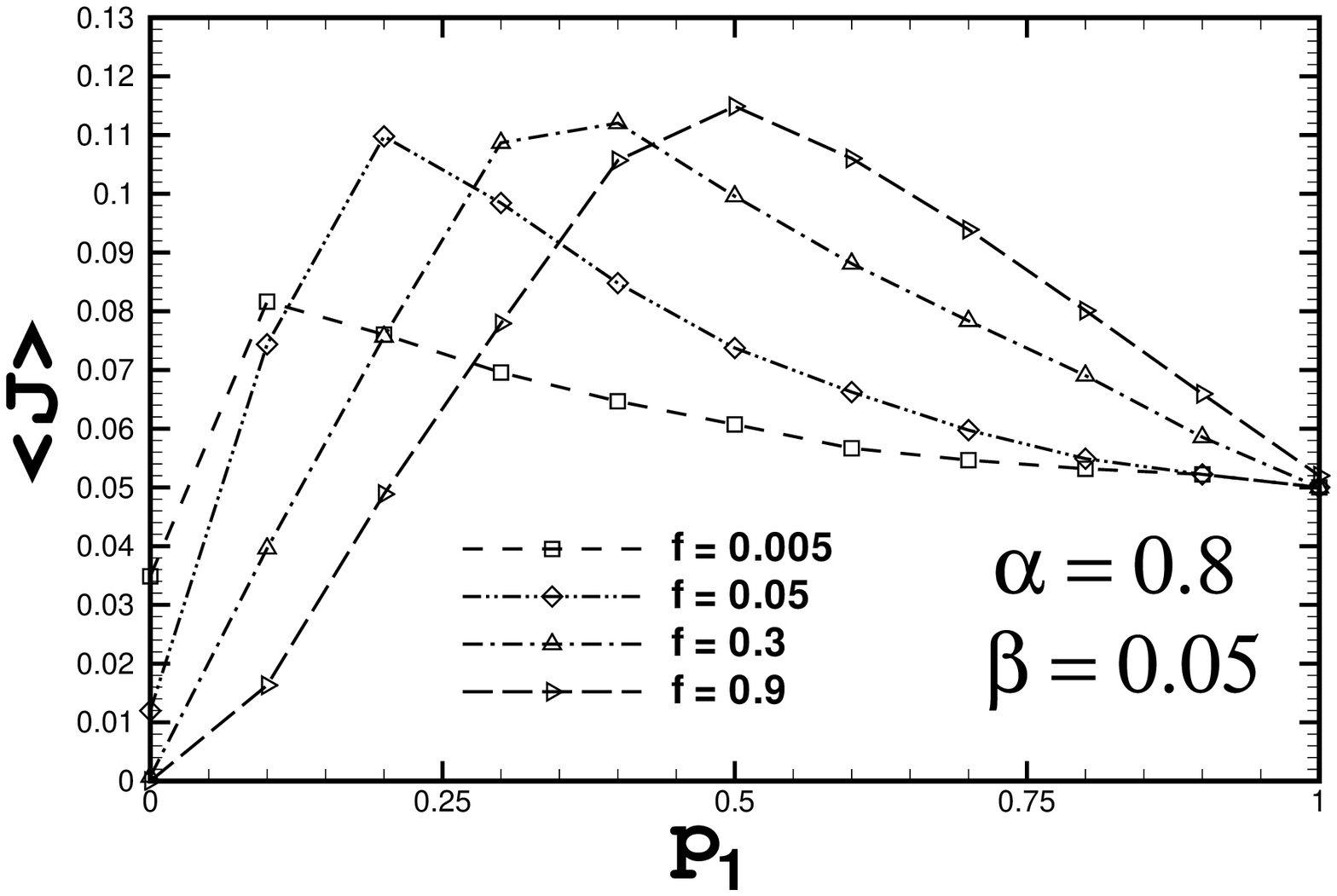}
\includegraphics[width=7cm]{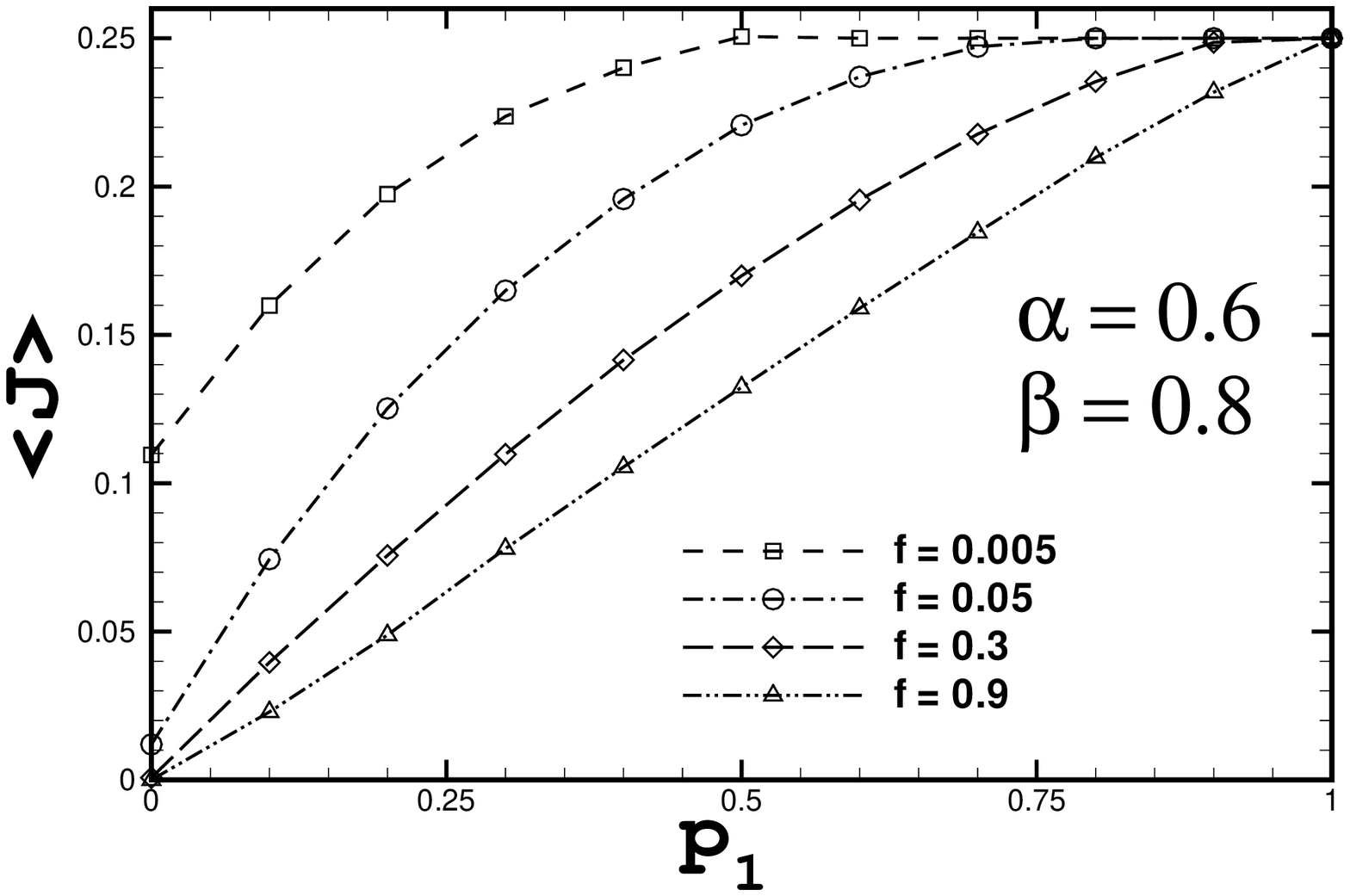}

\caption{ Figures~ 13-15: Current dependence on $p_1$ for various
values of $f$: low density phase $\alpha=0.05, \beta=0.8$ (top),
high density phase $\alpha=0.8, \beta=0.05$ (middle) and maximal
current phase $\alpha0.6, \beta=0.8$ (bottom). System size is
$L=300$. } \label{fig:bz2}
\end{figure}

When $\alpha$ is small and $\beta$ large, the effect of increasing
$p_1$ is to increase the current to its normal value
$\alpha(1-\alpha)$. For each $f$, $\langle J \rangle $ increases
with $p_1$ up to a certain value and then gets saturated. This
implies that below a certain strength, the defect strength is
incapable of affecting the current. This picture changes
dramatically when large $\alpha$ and small $\beta$ are taken into
account. In this case, $\langle J \rangle$ increases with $p_1$ up
to an $f$-dependent value and then starts decreasing. The maximum
current sustained by the system is considerable. While in the case
$\alpha=0.8$~$\beta=0.05$, the current for the normal ASEP is
$\beta(1-\beta) \sim 0.05$; here we observe that disorder can
remarkably enhance the current to almost a doubled value around
$0.12$.

\begin{figure}
\centering
\includegraphics[width=7cm]{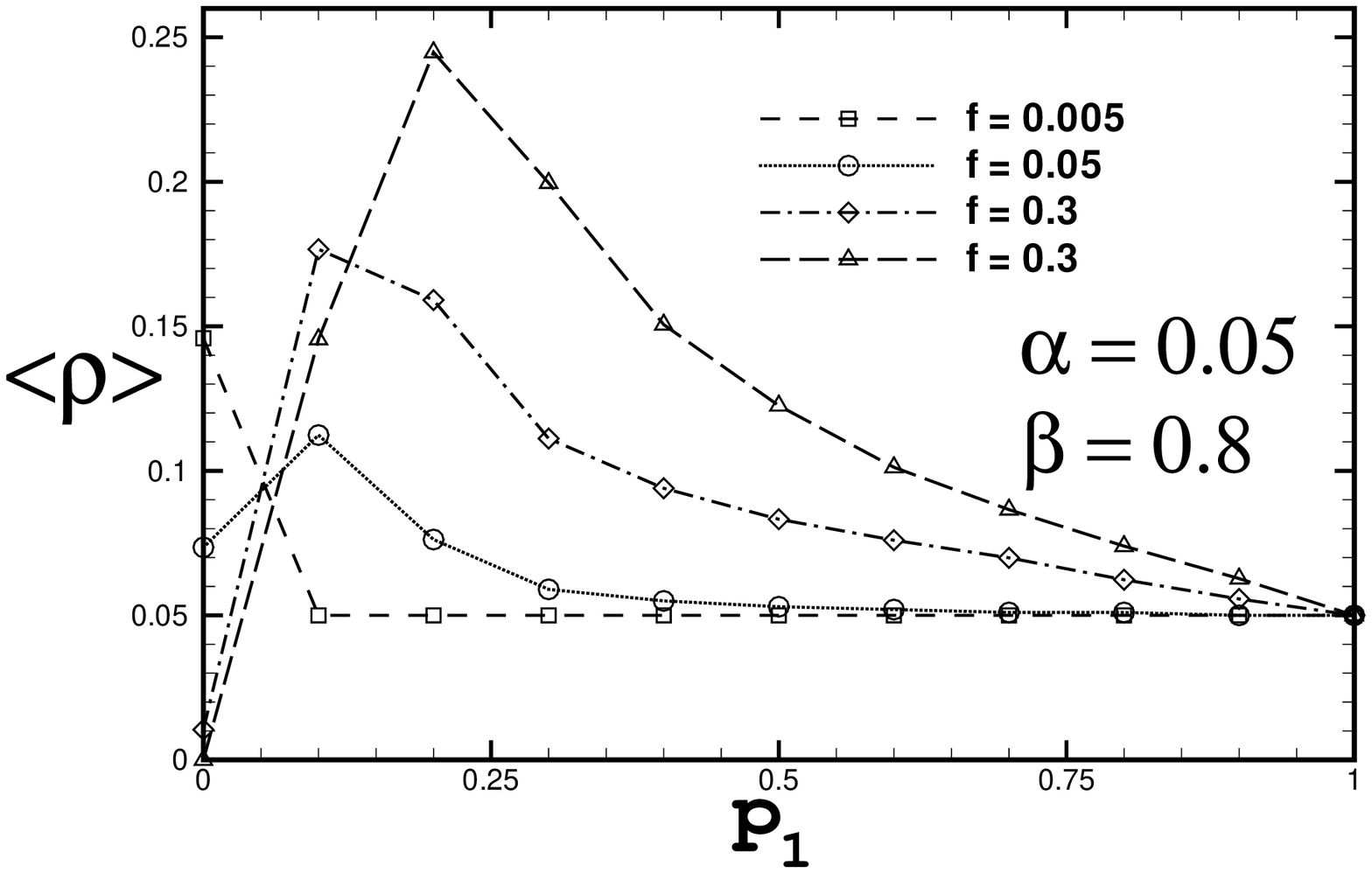}
\includegraphics[width=7cm]{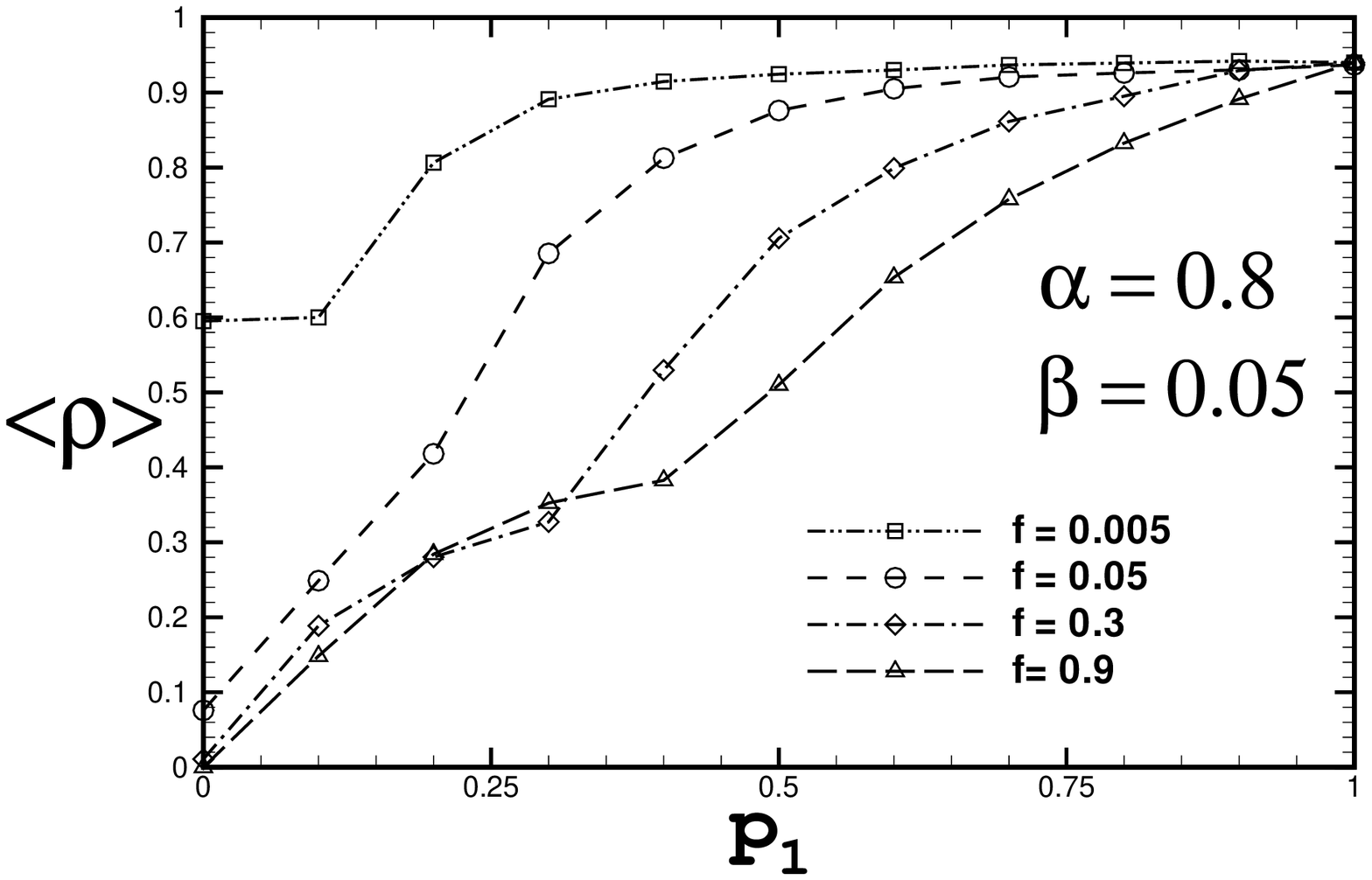}
\includegraphics[width=7cm]{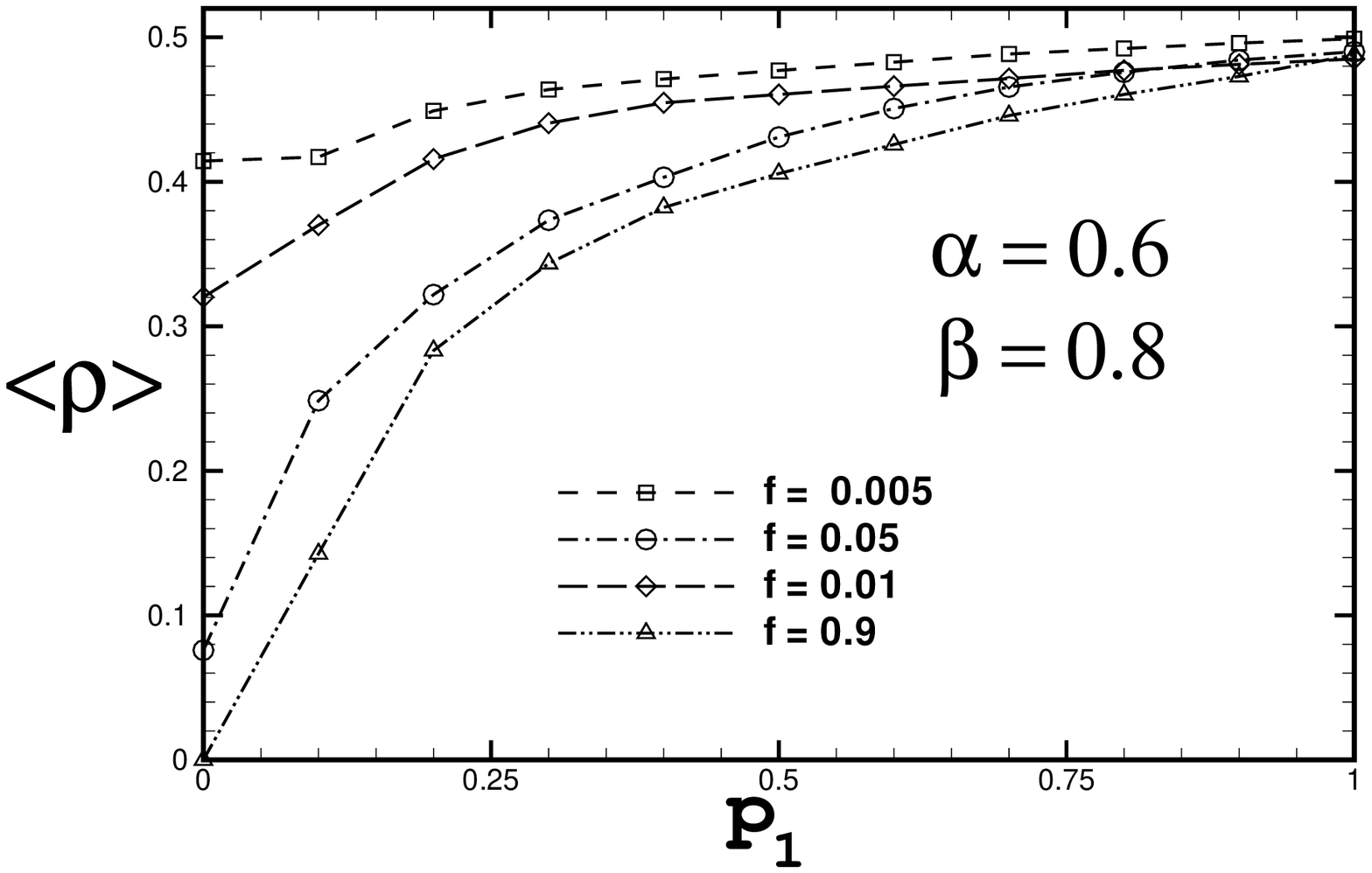}

\caption{ Figures~ 16-18: $\langle \rho \rangle$ vs $p_1$ for
various $f$: LD phase $\alpha=0.05, \beta=0.8$ (top), HD phase
$\alpha=0.8, \beta=0.05$ (middle) and MC phase $\alpha=0.6,
\beta=0.8$ (bottom). System size is $L=300$. } \label{fig:bz2}
\end{figure}

This type of constructive behaviour can be explained on the same
grounds as for figure (8). Qualitatively, the impurities do not
allow the overflow of particles into the system bulk which
otherwise would have led to congestion and current reduction. If
$p_1$ is small, the strength of defects is sufficient to block
the the inflow of particles and reduces the current. If $p_1$ is
high enough, large inflow $\alpha$ will dominate and $\langle J
\rangle$ is reduced. At an intermediate $p_1$ we have a maximal
current. Analogous to low $\alpha~$ high $\beta$, when both
$\alpha$ and $\beta$ are large, increasing $p_1$ leads to current
increments. If the density of defects is high, the current's
increase would be linear in $p_1$. For small $f$, the increase in
current is rather linear for small $p_1$ and afterwards becomes
more smooth. To deal in some depth, we next sketch the dependence
of $\langle \rho \rangle$ on the defect strength $p_1$. For small
$\alpha$ and large $\beta$ (see figure (16)), one interestingly
observes that if the defect concentration is relatively small
i.e., less than 0.02, the effect of decreasing the defect
strength (increasing $p_1$) is to reduce the density as
intuitively expected. Based on our previous arguments, defects
are more influential if their concentration is relatively small
\cite{chou}. Therefore, in the small concentration regime, the
weakening of the defects leads to a sharp decrease in the
density. Beyond a certain $p_1$, the further weakening of defects
does not affect $\langle \rho \rangle$. For defect concentration
$f$ above $0.02$, the behaviour of $\langle \rho \rangle$
undergoes a qualitative change. As observed in figure (16),
$\langle \rho \rangle$ increases up to a maximum and then starts
diminishing. The maximum value of $\langle \rho \rangle$ depends
on $f$ and ranges between $0.11$ and $0.25$. The reason is
twofold. First, for small input rate $\alpha$ and intermediate
concentration of defects, strong defects are still capable of
forming rather large high density regions behind them which
results in high $\langle \rho \rangle$. The second reason is due
to the blocking of the particles outflow. Although $\beta$ is
high, strong enough defects are able to reduce this high outflow
rate and effectively reduce it. Consequently, the bulk density
rises up throughout the bulk. Below a certain strength, the
defective sites, although their numbers are not so small, are not
only incapable of forming high density regions behind them but
also incapable of effectively reducing the output rate.
Therefore, $\langle \rho \rangle$ becomes decreasing. We now
consider the case where $\alpha$ is large but $\beta$ is small
(see fig. 17). Here the overall effect of decreasing the defect
strength is enhancement of $\langle \rho \rangle$. When the
defect strength is reduced, the particles can more easily enter
the chain and this leads to an increase in the bulk density. In
the limit of weak strength $p_1\rightarrow 1$, we recover the
normal value $\rho=1-\beta$. When both $\alpha$, and $\beta$ are
large corresponding to the MC phase in the normal ASEP, we still
observe that the effect of a reduction of the defect strength is
to enhance the density. Defects can drastically reduce the density
if their concentration and their strength are both large.
Otherwise, their influence is a slight reduction of the density.
We note the type of density increment is rather similar in
figures (17) and (18).

\section{Uniform distribution function}

So far our investigation has been restricted to the case where the
disorder strength was limited to only two values. In order to
obtain a complimentary insight into the nature of disorder effect,
it would be noteworthy to consider the case where the defect's
strength can be chosen from a continuous interval. For this
purpose, we consider the uniform distribution function for the
strength of defect. Here one has two parameters namely $a$ and $b$
which are the first and last points of the distribution interval.
Like the binary distribution, one can introduce two classes. In
the first class, $\langle p \rangle=\frac{a+b}{2}=1$ which
corresponds to the case having fast hopping sites greater than
unity. In the second class, $b=1$ while $a~(a<1)$ determines the
lower limit of the defect's value. Note that in the second class
$\langle p \rangle=\frac{1+a}{2} < 1$. We now exhibit the results
for the latter case i.e., $b=1$. The following diagram depicts
the dependence of $\langle J \rangle$ on the lower end of the
interval $a$. All diagrams which are shown next have been
obtained by Monte Carlo simulations. The number of disordered
samples over which the averaging have been performed is 1000, and
the system size is 200.

\begin{figure}
\centering
\includegraphics[width=7cm]{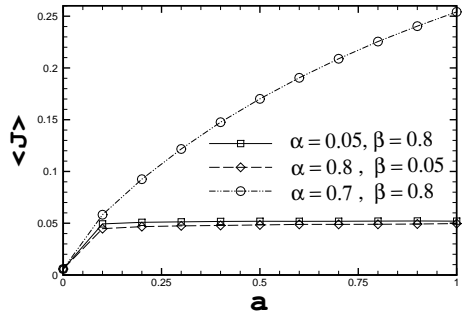}
\caption{ Figure 19~: Average current $\langle J \rangle$ versus $a$
for low $\alpha$ high $\beta$, high $\alpha$ low $\beta$ and high
$\alpha$ high $\beta$. $L=200$. } \label{fig:bz2}
\end{figure}

According to figure (19), in low $\alpha$-high $\beta$ and high
$\alpha$-low $\beta$ regimes, beyond $a=0.1$, the impurities do
not affect the current and the system can maintain the normal ASEP
values $\alpha(1-\alpha)$ and $\beta(1-\beta)$ respectively. In
contrast, for both $\alpha$ and $\beta$ larger than $0.5$,
$\langle J \rangle$ is a smooth increasing function of $a$. The
current reaches its normal ASEP value $0.25$ in the limit $a
\rightarrow 1 $. The behaviour of the average bulk density
$\langle \rho \rangle$ on $a$ is shown in figure (20).

\begin{figure}
\centering
\includegraphics[width=7cm]{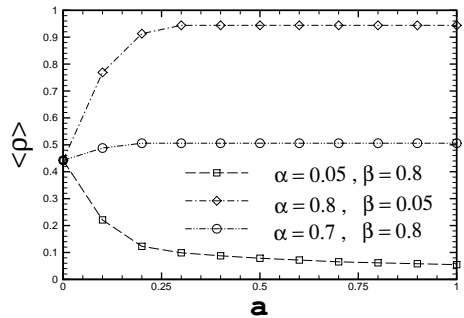}
\caption{ Figure 20~: Average density versus $a$ for low $\alpha$
high $\beta$, high $\alpha$ low $\beta$ and high $\alpha$ high
$\beta$. $L=200$. } \label{fig:bz2}
\end{figure}

Contrary to the current diagrams, here the density approaches the
normal ASEP value which depends on $\alpha$ and $\beta$. The high
$\alpha$-high $\beta$ regime has the weakest dependence on $a$ and
beyond $a=0.1$, $\langle \rho \rangle$ will be independent of
$a$. Figure (21) depicts the phase diagram in the case where the
average rate of hopping is unity: $\langle p \rangle=
\frac{a+b}{2}=1$ for various values of $a$. Analogous to the
binary distribution, the overall effect of disorder is to enlarge
the size of the MC phase and shrinkage of low and high density
phases, respectively.

\begin{figure}
\centering
\includegraphics[width=7cm]{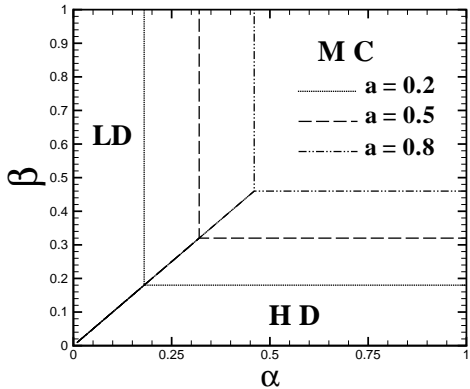}
\caption{ Figure 21~: Phase diagram of the disordered ASEP for a
uniform distribution of hopping rates. The distribution
characteristics are specified in the figures. All distributions have
unit mean but the variances are different. } \label{fig:bz2}
\end{figure}

Figure (22), exhibits the size dependence of the LD (HD) phase
$\delta$ in terms of the variance of the distribution functions for
both uniform and binary distribution functions. The mean of the
distribution functions is set to unity.

\begin{figure}
\centering
\includegraphics[width=7cm]{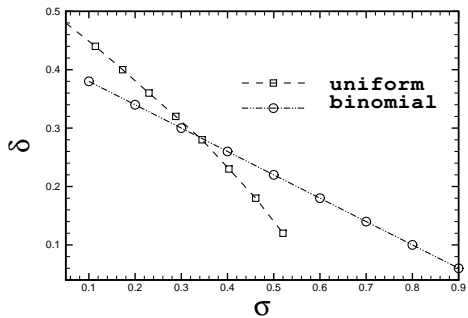}
\caption{ Figure 22~: Size of the low and the high density phases
versus the variance for uniform and binary distribution functions
($a=p_1$, $f=0.5$ and $\langle p \rangle =1$). $L=200$. }
\label{fig:bz2}
\end{figure}

For the uniform distribution, the size increment of the MC phase
shows a more rapid dependence on the variance of the distribution
function in comparison to the binary distribution. The reason is due
to the fact that in the uniform distribution, the frequency of
small-hopping sites close to the lower limit of the distribution
interval is more than those in the binary distribution.

\section{Summary and Concluding Remarks}

Let us now summarize what has been explored in this paper. We have
investigated the statistical characteristics of the asymmetric
simple exclusion process in the presence of spatially uncorrelated
quenched disorder in the hopping rates via extensive simulation
and numerics. Our findings cover two different distributions of
hopping rates: binary and uniform. The conventional three-phase
structure of the normal ASEP remains unchanged. Generically, the
disorder affects the phase diagram by enlarging the maximal
current phase, which in turn leads to squeezing the low and high
density phases. This is accompanied by an overall decrease
(increase) in the currents (densities). We have managed to
numerically solve the mean-field equations. Monte Carlo
simulations are in support of the mean field solutions. In brief,
the current exhibits a diminishing behaviour in terms of the
defect's concentration in the chain when the input rate is small
and the output rate is high. Analogously, it decreases when both
the input and output rates are relatively high. Unexpectedly, in
the case when the input rate is large and the output rate is
small, the current shows an increasing dependence versus the
defect's concentration. This demonstrates the nontrivial
interplay of spatial sitewise disorder with the drive. We have
also examined the properties of the ASEP under uniformly
distributed spatial disorder. Although the phase structure is
similar to that of with a binary distribution, we have identified
distinctive features between them. Our study has been limited to
disorder distribution functions with finite second moments. We
expect to observe substantial different types of behaviours for
those distributions having a long tail. Work along this line is in
progress.

\section{acknowledgement}

We wish to acknowledge the {\it Institute of Advanced Studies in
Basic Sciences} (IASBS) for proving us with the computational
facilities where the final stages of this work were carried out.
Fruitful discussion with Mustansir Barma is appreciated.

\bibliographystyle{unsrt}

\end{document}